%% file: main.tex
\documentclass[conference]{IEEEtran}
\usepackage[normalem]{ulem}
\usepackage{amsmath,amssymb,amsfonts}
\usepackage{algorithmic}
\usepackage{graphicx}
\usepackage{textcomp}
\usepackage{xcolor}
\def\BibTeX{{\rm B\kern-.05em{\sc i\kern-.025em b}\kern-.08em
    T\kern-.1667em\lower.7ex\hbox{E}\kern-.125emX}}

\usepackage{biblatex}
\usepackage{booktabs}

\usepackage{hyperref}
\usepackage{braket}
\usepackage{bm}
\usepackage{dsfont}
\usepackage{tikz}
\usetikzlibrary{decorations.pathreplacing,decorations.pathmorphing}
\hypersetup{
  colorlinks   = true, 
  urlcolor     = blue, 
  linkcolor    = blue, 
  citecolor   = green!60!black 
}
\usepackage{mathtools}

\makeatletter
\newcommand{\linebreakand}{%
  \end{@IEEEauthorhalign}
  \hfill\mbox{}\par
  \mbox{}\hfill\begin{@IEEEauthorhalign}
}
\makeatother
    
\begin{document}

\title{Sparse Simulation of VQE Circuits}

\author{
\IEEEauthorblockN{Damian S. Steiger}
\IEEEauthorblockA{\textit{Amazon Quantum Solutions Lab}\\
Zurich, Switzerland \\
dsteig@amazon.ch}
\and
\IEEEauthorblockN{Thomas Häner}
\IEEEauthorblockA{\textit{Amazon Quantum Solutions Lab}\\
Zurich, Switzerland \\
thaener@amazon.ch}
\linebreakand
\IEEEauthorblockN{Scott N. Genin}
\IEEEauthorblockA{\textit{OTI Lumionics Inc.} \\
Mississauga, Canada \\
scott.genin@otilumionics.com}
\and
\IEEEauthorblockN{Helmut G. Katzgraber}
\IEEEauthorblockA{\textit{Amazon Quantum Solutions Lab}\\
Seattle, USA \\
katzgrab@amazon.com}
}

\maketitle

\begin{abstract}
The Variational Quantum Eigensolver (VQE) is a promising algorithm for future Noisy Intermediate-Scale Quantum (NISQ) devices to simulate chemical systems. In this paper, we consider the classical simulation of the iterative Qubit Coupled Cluster (iQCC) ansatz.
To this end, we implement a multi-threaded sparse wave function simulator and simulate iQCC circuits with up to 80 qubits and 980 entanglers to compare our results to experimental values and previous approximate simulations.
In contrast to previous iQCC simulations, e.g., for computing the emission spectra of a phosphorescent emitting material, our approach features a variational guarantee, such that the resulting energies are true upper bounds on the exact energies. Additionally, our method is two orders of magnitude more memory efficient because it does not store the transformed Hamiltonians.
Our theoretical analysis also enables the construction of ans\"atze with a limited number of nonzero amplitudes, for which our simulator can obtain exact results.
This will allow one to generate complex benchmarking instances for future NISQ devices and simulators. 
\end{abstract}


\section{Introduction}

The Variational Quantum Eigensolver (VQE)~\cite{peruzzo2014variational} is an algorithm that can be run on Noisy Intermediate-Scale Quantum (NISQ)~\cite{preskill2018quantum} devices in order to solve problems in quantum chemistry and material science. The most promising problem sizes for VQE are between 50 and 100 spin orbitals (qubits). For smaller problems, there exist classical methods that are exact. For systems larger than that, the number of terms in the Hamiltonian and, as such, the number of required measurements becomes impractically large~\cite{liu2022prospects}. At such scales, quantum phase estimation (QPE) becomes the method of choice and a \emph{fault-tolerant} quantum computer is needed.
Although there do not exist any \emph{exact} classical methods that enable routine calculations in the range of 50-100 spin orbitals, approximate methods do exist. Arguably the most prominent are coupled cluster (CC) methods \cite{raghavachari1989fifth}. 
Unfortunately, this method has uncontrollable approximation errors. In contrast, VQE is variational and is guaranteed to return an upper bound on the true energy.

In this work, we consider scalable simulation of the VQE algorithm on classical computers to bridge the gap until quantum devices with the required specifications exist.

Exact simulations using a full-state vector approach would require storing $2^n$ complex amplitudes, which is only feasible up to approximately 45 qubits~\cite{haner2017simulator}. However, we observe that the quantum circuits implementing VQE are often expressed using exponentials of Pauli matrices, as is the case in, e.g., the qubit coupled cluster (QCC)~\cite{qcc2018} or the iterative qubit coupled cluster (iQCC) ansatz~\cite{iqcc2020}. These operators are (at most) two-sparse, and hence we opt for a sparse wave function representation~\cite{jaques2022leveraging} in this work.

We consider systems up to 80 qubits and 1100 entanglers,
and we carry out an analysis of the maximum number of nonzero elements that may occur in the wave function. We use the ansatz generated by iQCC \cite{largescaleiqcc2022} and find that, while we cannot store all nonzero amplitudes, a large fraction of amplitudes are negligibly small. We thus repeat our simulations with different cutoff parameters until the results converge. Moreover, we show that our results are comparable to iQCC energies and experimental values where available.

Compared to iQCC, our sparse simulation of VQE has two main advantages. First, it requires significantly less memory, and second, despite the approximation error due to the amplitude cutoff, our method produces a true upper bound on the ground-state energy as the Hamiltonian remains unchanged.

We note that our work does not include building the ansatz, i.e., evaluating the gradients of various possible operators using the parameter shift rule, and we did not optimize the parameters in the ansatz using our sparse simulator which allows us to investigate the effect of the sparse simulation cutoff \footnote{For each cutoff value we could further optimize the ansatz parameters, however, it would then be unclear if energy differences originate from improved parameters or from the cutoff in the sparse simulator.}. Instead, we rely on the results from iQCC. While parameter optimization would be possible using our simulator, we leave the implementation to future work.

\subsection{Related work}

Jaques and H\"aner~\cite{jaques2022leveraging} introduced a sparse wave function simulator and applied it to chemistry problems with up to 8 spin-orbitals, simulating the entire quantum phase estimation algorithm using a Trotter- and qubitization-based implementation.

Mullinax et al.~\cite{mullinax2023large} used a sparse simulator for VQE using the Unitary Coupled Cluster (UCC) ansatz. They demonstrated simulation capabilities up to 64 qubits, although they noted that this code limitation could easily be removed. They observed that the energy decreased as the number of nonzero elements stored in their simulator was increased to $10^5$ entries. The paper does not contain performance data for their simulator.

Morita et al.~\cite{morita2024simulator} used a distributed full-state vector simulator to simulate VQE circuits up to 32 qubits. This approach does not scale to sizes of 50-100 qubits, which are the most interesting problem sizes for future applications involving VQE.

\section{Methods}
In this section, we give a brief introduction to VQE before discussing the advantages and details of our sparse simulation method. Then, we derive an upper bound on the number of nonzero elements for a given VQE ansatz. We discuss the necessary quantum resources to run these circuits on future quantum devices and, finally, we present an overview of how iterative Qubit Coupled Cluster (iQCC) is evaluated on classical computers.

\subsection{Variational Quantum Eigensolver}
The Variational Quantum Eigensolver (VQE) algorithm aims to find the ground-state energy $E_0$ of a system described by a Hamiltonian $H$.
A parameterized wave function $\ket{\Psi(\bm\theta)}$ is constructed as follows:
\begin{equation}\label{eq:trial_wave function}
    \ket{\Psi (\bm\theta)} := V_m(\bm\theta) \ket{\Phi_0} = U_m(\theta_m) \dots U_1(\theta_1)\ket{\Phi_0},
\end{equation}
where $\ket{\Phi_0}$ is the initial Hartree-Fock state, i.e., a computational basis state with the lowest-energy orbitals filled with electrons, to which one applies a sequence of parameterized unitaries $U_k$ in order to prepare the so-called ansatz wave function or trial state $\ket{\Psi(\bm\theta)}$.

The variational principle states that the energy of this trial state is greater than or equal to the unknown ground state energy $E_0$:
\begin{equation}\label{eq:energy}
    E_0 \leq E_m(\bm\theta) = \bra{\Psi (\bm\theta)} H\ket{\Psi(\bm\theta)}.
\end{equation}
Finding an approximation of the ground state energy can be achieved by finding the parameters $\theta_k$ that minimize the energy $E_m(\bm\theta)$ of the trial state using a classical optimization procedure. Note that there is no efficient optimization algorithm for this problem in general~\cite{larocca2024review, kempe2006complexity}.

A useful feature of VQE is that the energy produced is an upper bound on the true ground-state energy. This stands in contrast to other classical approximation methods such as Coupled Cluster, where approximation errors are not guaranteed to obey the variational principle.

We consider the case where the unitaries $U_k$ are of the form
\begin{equation} \label{eq:pauli_exponentials}
    U_k = e^{i \theta_k P_k}, \; \mathrm{with\;} P_k \in
    \Pi_n:=\{\mathds{1}, X, Y, Z\}^{\otimes n} \mathrm{,\;} \theta_k \in \mathbb{R},
\end{equation}
where $n$ is the number of qubits and $P_k$ are multi-qubit Pauli operators constructed from single-qubit Pauli operators, given by
{\renewcommand{\arraystretch}{1.1}
\begin{center}
\begin{tabular}{ll}
    $\mathds{1}:= \begin{pmatrix} 1 & 0 \\ 0 & 1 \end{pmatrix}$,&$X:=\begin{pmatrix} 0 & 1 \\ 1 & 0 \end{pmatrix}$, \\  $Y:=\begin{pmatrix} 0 & -i \\ i & 0 \end{pmatrix}$, &$Z:=\begin{pmatrix} 1 & 0 \\ 0 & -1 \end{pmatrix}$.
\end{tabular}
\end{center}
}
This ansatz is called Qubit Coupled Cluster (QCC) in the literature. A similar form is obtained using the Unitary Coupled Cluster (UCC) ansatz after a Trotter-Suzuki expansion~\cite{trotter1959product, suzuki1976generalized}. We note that this ansatz is fully general because any unitary can be written as a product of Pauli exponentials.

\subsection{Sparse simulation of quantum circuits}
An $n$-qubit wave function $\ket{\Psi}$ can be represented as a vector of $2^n$ complex amplitudes $\alpha_i$
\begin{equation}
    \ket\Psi = \sum_{i=0}^{2^n-1} \alpha_i \ket{i},
\end{equation}
where $\ket{i}$ is computational basis state, i.e., $i\in \{ 0,1\}^n$ when viewed as a binary string that represents the integer $i$. Full state-vector simulators store this vector of amplitudes in its entirety. This means that, for example, as in Ref.~\cite{haner2017simulator}, 0.5 petabytes of RAM are needed to simulate 45 qubits.

Here, we make use of the state sparsity by only storing nonzero amplitudes as unordered key-value pairs ($i$, $\alpha_i$) in a hash map, as done in Ref.~\cite{jaques2022leveraging}.

The ansatz that we consider in this work consists of a sequence of Pauli exponentials, which are defined through a Pauli $P\in \Pi_n$ and a real parameter $\theta\in\mathds R$. Note that Pauli exponentials can be rewritten as
\begin{equation}\label{eq:sincosdecomp}
    e^{i\theta P} = \mathds 1\cos\theta + iP\sin\theta.
\end{equation}
Since $P$ is a Kronecker product of single-qubit Pauli gates $\Pi_1=\{\mathds 1,X,Y,Z\}$, which only have one nonzero entry per row/column, $P$ is one-sparse. As a result, $e^{i\theta P}$ is at most two-sparse~\cite{jaques2022leveraging}. In fact, it is one-sparse only if all Kronecker factors of $P$ are in $\{\mathds 1, Z\}$.

Although Pauli exponentials could be decomposed into one- and two-qubit gates, see Fig.~\ref{fig:circuit}, doing so would increase the number of nonzero elements in the hashmap during intermediate stages of the simulation. Moreover, this would increase the number of gates to simulate. Our simulator thus simulates the action of Pauli exponentials at a higher level of abstraction, which is also referred to as \textit{emulation}~\cite{haner2016high}.

Our simulator removes small amplitudes $\alpha_i$ if $\alpha_i^\dagger \alpha_i = | \alpha_i|^2$ is smaller than a predefined cutoff parameter. A renormalization step after each gate ensures that the wave function remains normalized.

At the end of the simulation, one measures the energy of the system described by the Hamiltonian $H$. The Hamiltonian is given as a linear combination of Pauli operators, i.e.,
\begin{equation}
    H = \sum_i h_i P_i,\,\mathrm{with\;} P_i \in \Pi_n,\,h_i \in \mathds R
\end{equation}
This form of a Hamiltonian for a molecular system can be obtained by transforming the second quantized fermionic Hamiltonian, for example, using the Jordan-Wigner \cite{Jordan1928,ortiz2001,Somma2002} or Bravyi-Kitaev transformation \cite{bravyi2002}.  
To measure the energy \eqref{eq:energy}, the simulator measures each term individually, i.e.,
\begin{equation} \label{eq:ind_pauli_measurement}
        E_m(\bm\theta) = \bra{\Psi (\bm\theta)} H\ket{\Psi(\bm\theta)} = \sum_i h_i \underbrace{\bra{\Psi (\bm\theta)} P_i \ket{\Psi(\bm\theta)}}_{=: \langle P_i\rangle_{\ket{\Psi}}} .
\end{equation}
We may write the probability of observing a $\pm 1$ outcome when measuring the $i$-th Pauli as $p(m_i=\pm 1)=\|\frac 12(\mathds 1\pm P_i)\ket{\Psi}\|^2$ and the corresponding expectation value as
\begin{equation}\label{eq:pauliexpectation}
    \langle P_i\rangle_{\ket\Psi} =2\cdot p(m_i=+1)-1.
\end{equation}

\subsection{Analyzing sparsity}\label{sec:sparsity_analysis}
We now derive an upper bound on the number of nonzero elements in any trial wave function \eqref{eq:trial_wave function} prepared by Pauli exponentials with Pauli operators $P_1,...,P_m$ as in \eqref{eq:pauli_exponentials}. 

The initial state is the Hartree-Fock state, which is a computational basis state denoted by $\ket x$. Applying a Pauli exponential to this computational basis state yields two summands~\cite{jaques2022leveraging} (see \eqref{eq:sincosdecomp}),
\begin{equation}\label{eq:sincosstatedecomp}
    \cos\theta \cdot \ket{x} + i\sin\theta \cdot P_i\ket{x}.
\end{equation}
The second summand involves applying a Pauli $P_i\in\Pi_n$ to a computational basis state $\ket{x}$. For a given $P_i\in\Pi_n$, we denote the corresponding XY-type mask as $P_i|_{XY}$, where $P_i|_{XY}$ is a bit-vector with a 1 in position $i$ if and only if the $i$-th Kronecker factor of $P_i$ is a Pauli $X$ or $Y$. Using this notation, we have
\begin{equation} \label{eq:applypauli}
    P_i \ket{x} = c \ket{x\oplus P_i|_{XY}}, \quad c\in \mathds C.
\end{equation}
The $X$- and $Y$-factors in $P_i$ flip the corresponding bit of $x$. Let us consider the final amplitude $\alpha_y\in\mathds C$ corresponding to a computational basis state $\ket y$ after applying a sequence of $m$ Pauli exponentials \eqref{eq:trial_wave function} to a computational basis state $\ket{x}$ (e.g., the Hartree-Fock state):
\begin{equation}
    \alpha_y=\bra{y} e^{i\theta_mP_m}\cdots e^{i\theta_1P_1}\ket{x}.
\end{equation}
We introduce a binary decision variable for each of the $m$ Pauli exponentials that specifies whether we consider the first summand (with the $\cos\theta$ prefactor) or the second summand (with the $i\sin\theta$ prefactor) in \eqref{eq:sincosstatedecomp}. We denote these decision variables as $p_1,...,p_m$, where $p_i\in\{0,1\}$ and $p_i = 1$ means that we consider the second term, that is, $iP\sin\theta$ in \eqref{eq:sincosdecomp}. We can hence rewrite the equation above as
\begin{align}\label{eq:single_amplitude}
    \alpha_y=\sum_{p_1,...,p_m\in\{0,1\}^m}\bra{y} & (c_m{\overline{p_m}} + s_m{p_m}P_m)\cdots \\ \nonumber 
    & (c_1{\overline{p_1}} + s_1{p_1}P_1)\ket{x},
\end{align}
where $c_k:=\cos\theta_k$, $s_k:=i\sin\theta_k$ and $\overline p=1-p$ for $p\in\{0,1\}$. From \eqref{eq:applypauli}, it is apparent that the terms in the sum above can be nonzero only if
\begin{equation}\label{eq:gf2lin}
    x\oplus y = \bigoplus_{i=1}^m p_i P_i|_{XY} .
\end{equation}
From this expression, it follows that one may derive an upper bound on the number of nonzero elements in the trial wave function based on the rank of the matrix $P|_{XY}$, whose columns are given by the $P_i|_{XY}$ vectors. Specifically, there are at most $2^{\operatorname{rank}(P|_{XY})}$ nonzero elements in $\ket{\Psi (\bm\theta)}$. We note that this upper bound is tight for some choices of parameters $\theta_i$.

As a side note, if one is interested in calculating only an individual amplitude $\alpha_y$, then it is not necessary to compute all $2^m$ terms in the sum~\eqref{eq:single_amplitude}. Instead, it is sufficient to sum over all solutions $\bm p = (p_1,...,p_m) \in \{0, 1\}^m$ of the linear system of equations \eqref{eq:gf2lin}.

\subsection{Quantum Resources}
Exponentials of multi-qubit Pauli operators, $U(\theta)=\exp(i\theta P)$, can be implemented on a quantum computer using the decomposition shown in Fig.~\ref{fig:circuit}, where one applies a single-qubit basis change for each $X$ and $Y$ factor in $P$ using
\begin{align}
    H &:= \frac{1}{\sqrt{2}}\begin{pmatrix} 1 & 1 \\ 1 & -1 \end{pmatrix} \\ \nonumber
    R_x(\theta) &:= e^{-i\theta X/2} = \begin{pmatrix} \cos{\frac{\theta}{2}} & -i\sin{\frac{\theta}{2}} \\ -i\sin{\frac{\theta}{2}} & \cos{\frac{\theta}{2}}, \end{pmatrix},
\end{align}
followed by a parity calculation using CNOTs, the rotation by $\theta$, and a subsequent uncomputation of the parity and the initial basis change~\cite{whitfield2011simulation}.

A future NISQ device would measure each Pauli $P_i$ individually as in \eqref{eq:ind_pauli_measurement}. In practice, there are multiple methods to reduce the number of measurements of the set of Pauli operators $P_i$. The original VQE algorithm running on a NISQ quantum computer creates a set of single-qubit operators that diagonalize each $P_i$ individually into the Pauli $Z$ basis. Expectation values may then be computed via repeated measurements. To reduce the number of measurements, multiple methods have been proposed, ranging from grouping commuting sets~\cite{jena2021physrevA,minium_clique_measurement2020,allcompatibleopterators2020} to using the fermionic algebra of the original Hamiltonian~\cite{Choi2023fluidfermionic}.

The required quantum resources for all systems can be found in subsection~\ref{sec:sub_quantum_resources}.

\begin{figure*}[ht]
    \begin{center}
        \input{time_evolution.tikz}
    \end{center}
    \caption{Decomposition of multi-qubit Pauli exponential into a quantum circuits using only single-qubit and two-qubit CNOT gates~\cite{whitfield2011simulation}. Gates between ${\color{blue}\raisebox{.5pt}{\textcircled{\raisebox{-.4pt}{\scriptsize{0}}}}} \rightarrow {\color{blue}\raisebox{.5pt}{\textcircled{\raisebox{-.4pt}{\scriptsize{1}}}}}$ are a basis changes on each qubit to transform $X$ and $Y$ to Pauli $Z$, gates between ${\color{blue}\raisebox{.5pt}{\textcircled{\raisebox{-.4pt}{\scriptsize{1}}}}} \rightarrow {\color{blue}\raisebox{.5pt}{\textcircled{\raisebox{-.4pt}{\scriptsize{4}}}}}$ implement ${\rm exp}(-i\theta Z_1 Z_2 Z_3)$, gates between ${\color{blue}\raisebox{.5pt}{\textcircled{\raisebox{-.4pt}{\scriptsize{4}}}}} \rightarrow {\color{blue}\raisebox{.5pt}{\textcircled{\raisebox{-.4pt}{\scriptsize{5}}}}}$ undo the basis change.} 
    \label{fig:circuit}
\end{figure*}

\subsection{iQCC Method}
The iterative Qubit Coupled Cluster (iQCC) method~\cite{iqcc2020} builds a trial wave function as in \eqref{eq:trial_wave function} using a sequence of Qubit Coupled Cluster (QCC) entanglers of the form 
\begin{equation} \label{eq:qcc_entanglers}
    U_k(\beta_k) := \exp(-i\beta_k P_k /2) \quad \mathrm{with}\;\beta_k \in \mathbb{R}.
\end{equation}
Note that these differ from $U_k(\theta)$ in \eqref{eq:pauli_exponentials} in the definition of the parameter, i.e., $\theta_k = -\beta_k/2$. The full trial wave function is then given by
\begin{equation}\label{eq:iqcc_trial_wave function}
    \ket{\Psi (\bm\theta)} = V_m(\bm\beta) \ket{\Phi_0} = U_m(\beta_m) \dots U_1(\beta_1)\ket{\Phi_0}
\end{equation}

This is called the QCC ansatz, or iQCC ansatz if it is built iteratively~\cite{qcc2018, iqcc2020}. Its individual entanglers, i.e., the Pauli exponentials $\exp(-i\beta_k P_k/2)$, are constructed from the molecular Hamiltonian such that the energy gradient with respect to the parameter $\beta_k$ at $\beta_k=0$ is large. The product form of this ansatz enables an exact implementation on a quantum computer as each QCC entangler can be implemented as shown in Fig.~\ref{fig:circuit}. In contrast, the also popular unitary coupled cluster (UCC) ansatz is constructed from a fixed set of fermionic excitation and annihilation operators independent of the specific molecule~\cite{romero2018strategies}.

The iQCC ansatz could be evaluated on a future quantum computer. However, it can also be (approximately) evaluated using classical computers~\cite{largescaleiqcc2022} already today. As done in previous studies, the iQCC procedure starts with $P_1$ and optimizes the value of $\beta_1$ to minimizes the energy
\begin{equation}
   \bra{\Phi_0}U_1(\beta_1)^\dagger H_0 U_1(\beta_1)\ket{\Phi_0},
\end{equation}
where we labeled the original Hamiltonian $H$ of the system as $H_0$.
After the optimization of the parameter $\beta_1$, the Hamiltonian is transformed using the best value found for $\beta_1$, which we denote by $\beta_1^*$, resulting in
\begin{align}\label{eq:hamtrafo}
   H_1 &:=  U_1(\beta_1^*)^\dagger H_0 U_1(\beta_1^*) \\ \nonumber
   &= H_0 -i \frac{\sin(\beta_1^*)}{2}[H_0,P_1] \\ \nonumber 
   &+ \frac{1}{2} \left(1-\cos\left(\beta_1^*\right)\right)(P_1H_0P_1 - H_0).
\end{align}
Because this new Hamiltonian $H_1$ is a similarity-transformed version of $H_0$, it has the same eigenvalues by construction. The iterative procedure continues by searching for a new unitary $U_2(\beta_2)$ that lowers the energy for $H_1$ and so on.

A limiting factor is that the number of terms in this new Hamiltonian initially grows exponentially with the number of iterations due to the transformation in Eq.~\eqref{eq:hamtrafo}, before eventually plateauing~\cite{largescaleiqcc2022}. There are techniques to reduce the growth of the Hamiltonian, but these come with various trade-offs and have not been tested on systems with more than 36 qubits~\cite{growthreduction}.

For the purposes of this publication, the Direct Interaction Space (DIS) method for selecting QCC entanglers is used, because it has the most consistent energy minimization and is well studied in the literature~\cite{qcc2018, iqcc2020, largescaleiqcc2022}. 

It is important to note that the numerical approximations introduced by transforming the Hamiltonian at each step might result in the final Hamiltonian being not isospectral to the original Hamiltonian $H$. As a result, the optimized parameters $\beta^*_i$ evaluated on a quantum computer or using our sparse simulator may not produce a state with the exact same energy as the one returned by iQCC. However, because the sparse simulator measures the energy of the ansatz using the original Hamiltonian, our method returns a true upper bound on the exact energy of the system, i.e., it satisfies the variational principle for any choice of amplitude cutoff. Depending on the concrete use case (e.g. solving the chemistry problem classically as opposed to simulating the circuit accurately), one may want to use an amplitude cutoff that results in the lowest energy, as opposed to reducing the cutoff until the energy converges (see Fig.~\ref{fig:cutoff}).

\section{Experimental setup}
We run our sparse wave function simulator on a \emph{r7i.metal-48xl} Amazon EC2 instance~\cite{awsec2r7i}. These instances have two sockets with Intel(R) Xeon(R) Platinum 8488C CPUs, each with 48 cores and two threads per core.

\begin{figure}[ht]
\centerline{\includegraphics{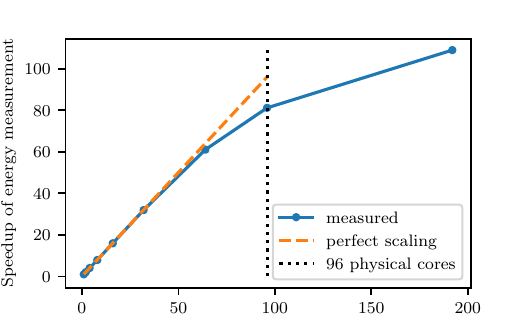}}
\caption{Strong scaling behavior of the energy measurement for the smaller N$_2$ dimer. Amplitudes were removed if $|\alpha_i|^2 < 10^{-11}$. Going from 96 threads to 192 threads, i.e., using hyperthreading with 2 threads per core, improves performance by 26\%.}
\label{fig:strong_scaling}
\end{figure}

The sparse simulator is implemented in C\texttt{++} using the hash map from Ref.~\cite{skarupke2018}. The application of gates is implemented in a single-threaded fashion, while the energy measurement~\eqref{eq:pauliexpectation} is executed in parallel using OpenMP, because this is the most computationally expensive part of the simulation. For example, the single-threaded simulation time for our smaller system of the N$_2$ molecule is 86 seconds, whereas the single-threaded energy measurement takes 7070 seconds. We have included a strong scaling plot in Fig.~\ref{fig:strong_scaling}.

We use a modified version of GAMESS to obtain the one- and two-electron integrals~\cite{GAMESS} using the frozen core approximation, where molecular orbitals outside of the complete active space (CAS) are not acted upon. The integrals are then transformed into qubit operators using a Jordan-Wigner transformation.

\section{Results}
\begin{figure}[ht]
    \begin{center}
    \includegraphics[width=0.8\linewidth]{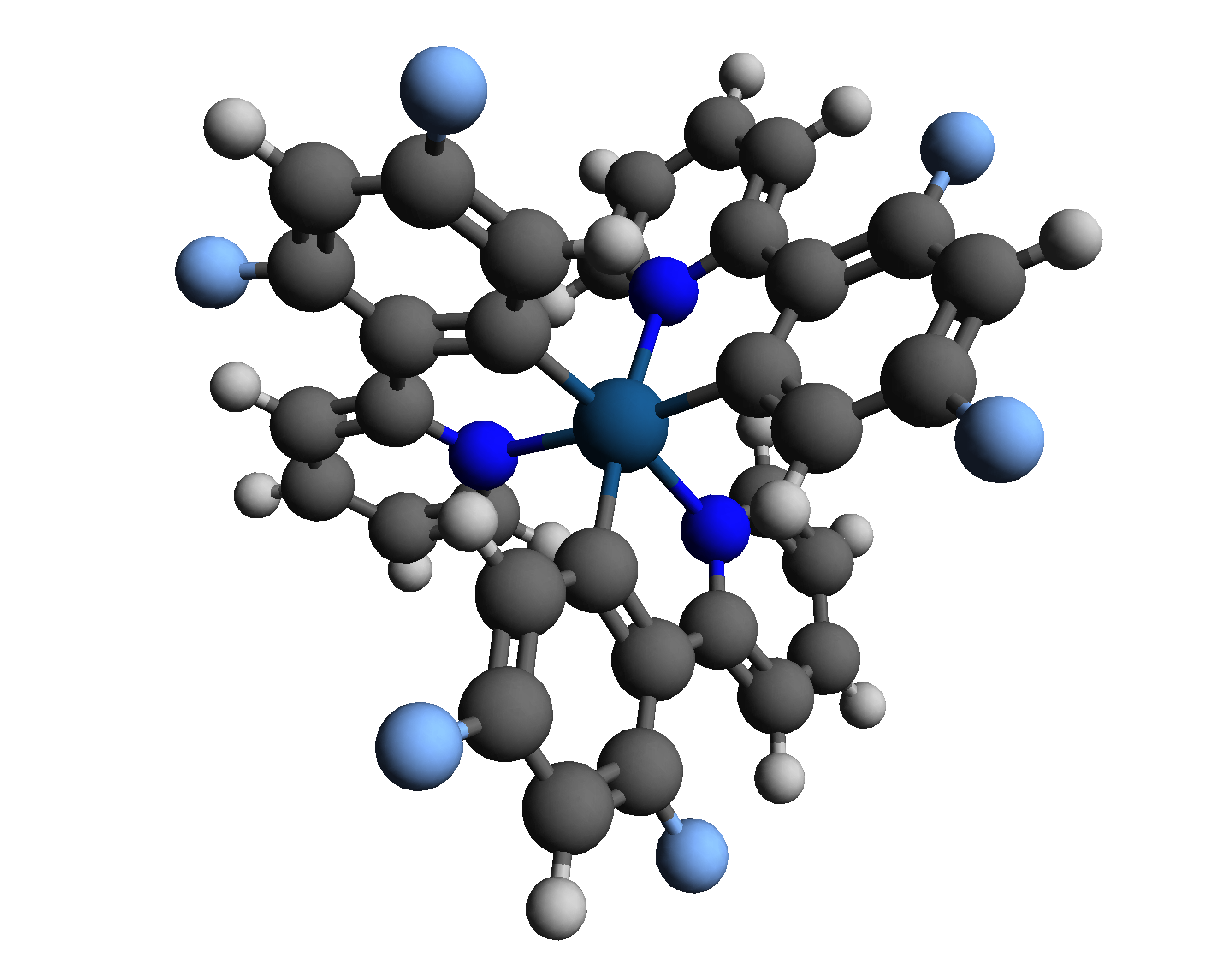}
    \end{center}
    \caption{Iridium~(III) phosphorescent emitting material, Ir(F$_2$ppy)$_3$, which we use as a benchmark. Image generated using \cite{hanwell2012avogadro}.}
    \label{fig:irf2ppy3}
\end{figure}

To test and benchmark our sparse simulator, we consider two systems: diatomic nitrogen, N$_2$, and a Iridium~(III)
phosphorescent emitting material, Ir(F$_2$ppy)$_3$, see Fig.~\ref{fig:irf2ppy3}. 

Diatomic nitrogen, N$_2$, has one of the strongest bonds in chemistry~\cite{kalescky2013identification} and is a well-known benchmark for computational quantum chemistry~\cite{chan2004state}. 

Phosphorescent emitting materials are used to build organic light-emitting diode (OLED) screens for modern electronic devices. Phosphorescent emitting materials are important because they can have a quantum efficiency of 100\% since a photon can be emitted from the triplet state, compared to the quantum efficiency of fluorescent emitters that have a maximum efficiency of 25\%. Phosphorescent materials are typically more difficult to synthesize, because they are organo-metallic complexes and thus improvements in simulation accuracy can reduce the cost to design new OLED materials. We chose, Ir(F$_2$ppy)$_3$, because experimental data are available on the light-emitting spectra. This allows us to validate our simulation results against experimental data by computing the energy difference between the involved singlet (S$_0$) and triplet state (T$_1$). 

The input Hamiltonians and the outputs of the iQCC simulations (entanglers, energies, and optimized amplitudes) are available in Ref.~\cite{data}.

We note that an exact single-point energy calculation, that is, full configuration interaction (FCI), is infeasible for this system. Note that, compared to the exact energies, an energy difference of 1kcal/mol or $0.0016$ Hartree is considered to be of chemical accuracy~\cite{pople1999nobel}.

\subsection{Nitrogen Dimer Results}
As a small test system, we chose diatomic nitrogen N$_2$. We prepared the Hamiltonian of N$_2$ with a bond distance of $R=1.09$ Angstroms and the cc-pvdz basis set. 
This resulted in a CAS(12,28), i.e., 12 electrons in 28 molecular orbitals, which corresponds to 56 qubits.
The results for different cutoff parameters are shown in Table~\ref{tab:n2_results_opt_amp}. The values change by less than $0.2$mHa over a cutoff range from $10^{-11}$ to $10^{-14}$. For the smallest cutoff, there is a difference of $3.5$mHa compared to the iQCC results. This difference is most likely due to numerical approximations in iQCC, which removes tiny Hamiltonian terms after each iQCC step. Specifically, in this case, iQCC removed terms with coefficients less than $5.0 \cdot 10^{-7}$. Consequently, the similarity-transformed Hamiltonian $H_{69}$ and the starting Hamiltonian $H_0$ are not perfectly isospectral. In contrast, our sparse simulator measures the original Hamiltonian $H_0$. Moreover, all energies in Table~\ref{tab:n2_results_opt_amp} are true upper bounds, in particular $-109.226307$Ha at a cutoff of $5\cdot 10^{-7}$, which is very close to the iQCC energy of $-109.226315$Ha. 

\begin{table*}[t]
    \centering
    \begin{tabular}{cclrrrr}
        \toprule
        \#iQCC steps & cutoff & energy [Ha] & $\Delta$ [mHa] & sim. time [s] & meas. time [s] & \#elements \\\midrule
        69 & 1e-5  & -109.209798 & 16.687 & 0.06 & 0.22 & 836 \\
        69 & 5e-6  & -109.225389 & 1.096  & 0.07 & 0.25 & 1258 \\
        69 & 1e-6  & -109.226188 & 0.297 & 0.08 & 0.26 & 1487 \\
        69 & 5e-7  & -109.226307 & 0.178 & 0.1 & 0.2 & 1820 \\
        69 & 1e-7  & -109.226405 & 0.080 & 0.4 & 0.4 & 4555 \\
        69 & 5e-8  & \textbf{-109.226485} & \boldmath$\coloneq 0$ & 0.7 & 0.5 & 8072 \\
        69 & 1e-8  & -109.225527 & 0.958 & 2.5 & 1.3 & 30131 \\
        69 & 5e-9  & -109.224981 & 1.504& 4.3 & 2.2 & 52434 \\
        69 & 1e-9  & -109.223862 & 2.623 & 13 & 9  & 164881 \\
        69 & 5e-10 & -109.223537 & 2.948 & 17 & 14 & 244601 \\
        69 & 1e-10 & -109.223168 & 3.317 & 31 & 30 & 500224 \\
        69 & 5e-11 & -109.223095 & 3.390& 40 & 38 & 638665 \\
        69 & 1e-11 & -109.222999 & 3.486 & 87 & 65 & 1088320 \\
        69 & 1e-12 & -109.222892 & 3.593 & 484 & 440 & 4022955 \\
        69 & 1e-13 & -109.222834 & 3.651 & 2290 & 3135 & 17037969 \\
        69 & 1e-14 & -109.222811 & 3.674 & 6204 & 10661 & 53607413 \\
        \bottomrule \noalign{\vskip 1mm} 
    \end{tabular}
    \caption{Sparse simulator results for running N$_2$ with different amplitude cutoffs (lowest energy in bold and $\Delta$ is the difference to that energy). Amplitudes $\alpha_i$ are kept if $|\alpha_i|^2 \geq \text{cutoff}$. We used the optimized parameters $\beta^*$ from iQCC. Each iQCC step contains 20 QCC entanglers of the form \eqref{eq:qcc_entanglers}. In comparison the energy obtained using the iQCC algorithm on a classical computer resulted in an energy of $-109.226315$Ha. 
    }
    \label{tab:n2_results_opt_amp}
\end{table*}

\subsection{Iridium complex Results}
We chose Ir(F$_2$ppy)$_3$ as a model example for a well-studied phosphorescent emitting material, for which both simulation data using iQCC~\cite{largescaleiqcc2022} and well-documented experimental data~\cite{thompsonexperiment} are available.

We used the T$_1$- and S$_0$-optimized geometries from Ref.~\cite{largescaleiqcc2022} and the Hamiltonians were prepared using either the LANL2DZ-ECP or the SBKJC-ECP basis set on the Ir center and the 6-31G(d) basis set on all other atoms using a modified version of GAMESS~\cite{GAMESS}. This results in a CAS(40,40) for all Ir(F$_2$ppy)$_3$ calculations, i.e., 40 electrons in 40 molecular orbitals, which corresponds to 80 qubits.

\begin{figure}[ht]
    \includegraphics[width=\linewidth]{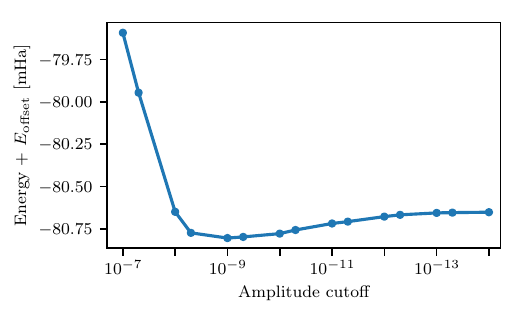}
    \caption{Energy obtained using the sparse simulator for the S$_0$ state of Ir(F$_2$ppy)$_3$ using the  LANL2DZ-ECP basis set with different amplitude cutoffs ranging from $10^{-7}$ to $10^{-14}$. $E_{\mathrm{offset}} \equiv 2124$Ha is a constant offset for easier readability. Complete raw data for a wider range of cutoffs is in Table~\ref{tab:s0_results_opt_amp}.}
    \label{fig:cutoff}
\end{figure}

We used the LANL2DZ-ECP basis set on the Ir centre and the T$_1$-optimized geometry to calculate both the T$_1$ and S$_0$ states, as this is the most representative of the phosphorescent process and allows us to compare against experimental values of the T$_1$-S$_0$ emission spectra. The results of the sparse simulator for different cutoff values are listed in Tables~\ref{tab:s0_results_opt_amp} and \ref{tab:t1_results_opt_amp} and in Fig.~\ref{fig:cutoff}.

We aim for our energy results to be converged to within less than $1$mHa and the energy is given as a sum over $3.7$ million terms of the Hamiltonian.

Starting at a cutoff of $10^{-9}$, we observe a convergence in Fig.~\ref{fig:cutoff}. The difference between the iQCC energies and ours are less than $0.5$mHa for both S$_0$ and T$_1$ states, which is well within chemical accuracy. Note that iQCC simulations were performed using spin and number constraints~\cite{constrainedVQE2019}. Here, we report the iQCC energies without the penalty terms used for the constraints.

The calculated T$_1$ to S$_0$ transition for Ir(F$_2$ppy)$_3$ after $49$ iQCC iterations is $2.676$eV using the iQCC evaluation method, $2.664e$V using the sparse simulator with a cutoff of $10^{-13}$, and    
$2.805$eV using iQCC with Epstein-Nesbit pertubation theory corrections (iQCC+PT) introduced in \cite{pt_correction2021}. Note that iQCC+PT is a non-variational energy correction to the variational iQCC energy. All three results are within $0.1$eV of the experimentally-reported result at $77$K of $2.73$eV~\cite{thompsonexperiment}.

\noindent\textbf{Memory requirements.} 
The transformed iQCC Hamiltonian after $49$ iQCC steps has $3\,030\,709\,562$ terms for the S$_0$ state and $3\,513\,047\,499$ terms for the T$_1$ state. The Pauli operators, $P_i$, are stored using two 128-bit integers and a double for the coefficient $h_i$ which results in 121GB of memory for the S$_0$ state and 141GB of memory for the T$_1$ state. We neglect the storage required of iQCC for transformed observables which are necessary for penalty terms. 

In contrast, the sparse simulator needs to store the nonzero elements of the wave function in a hash map, in addition to the original Hamiltonian, which has $3\,775\,249$ terms for the S$_0$ state and $3\,774\,557$ terms for the T$_1$ state, i.e., 151MB for the Hamiltonian.
The required memory for the hash map can be estimated using the following properties: a 128-bit integer is used for state labels in the hash map, a complex double for the corresponding amplitude, 1 byte overhead per hash map entry, and the hash map has a maximum load factor of $0.9375$. Moreover, the hash map doubles in size each time this load factor is reached, starting at 10 entries~\cite{hashmap_overhead}. Using the results from our simulation with a cutoff of $10^{-12}$, this results in a hash map size of $1.38$GB to store the sparse wave function for the S$_0$ or T$_1$ state. The memory savings compared to iQCC thus amount to approximately two orders of magnitude for this specific molecule. 

If the goal is not to accurately simulate the circuit, but to obtain a low energy upper bound, the memory savings can be even more substantial. For example, for the $S_0$ state, we find that our simulations agree best with iQCC at an amplitude cutoff of $10^{-9}$, where the number of elements in the hashmap is $20629402/354689\approx 58$ times smaller than at a cutoff of $10^{-12}$.

We note that in our code pipeline we had additional memory overheads that are not accounted for here. For example, we processed the data in Python and called the C++ simulator from there, which results in a copy of some of the data.

\begin{figure}[ht]
    \includegraphics[width=\linewidth]{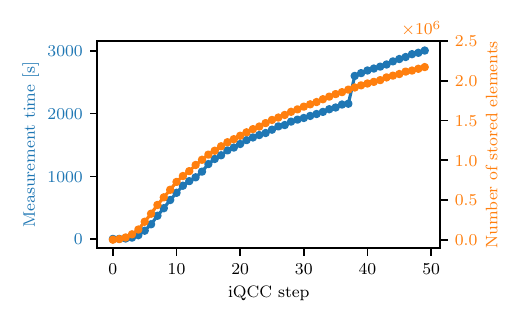}
    \caption{Measurement time and number of nonzero elements stored in the sparse simulator for the S0 state of Ir(F$_2$ppy)$_3$ using the LANL2DZ-ECP basis set with an amplitude cutoff of $10^{-11}$ as a function of the iQCC steps included in the simulation. Each iQCC step contains 20 QCC entanglers. The jump in measurement time after iQCC step 37 occurs because the sparse state no longer fits into the L3 cache of our CPU.}
    \label{fig:measurement_iqcc_steps}
\end{figure}

To determine whether the simulation time or the measurement time of the sparse simulator would be affected by the basis set, we also calculated the S$_0$ state in the S$_0$ geometry using the SBKJC-ECP basis set on the Ir centre, see Table~\ref{tab:s0_sbkjc_results_opt_amp}.
We show the measurement time as a function of the ansatz length in Fig.~\ref{fig:measurement_iqcc_steps}.

\begin{table*}[t]
    \centering
    \begin{tabular}{cclrrrr}
        \toprule
        \#iQCC steps & cutoff & energy [Ha] & $\Delta$ [mHa] & sim. time [s] & meas. time [s] & \#elements \\\midrule
        49 & 5e-5  & -2124.067435 & 13.369 & 0.03 & 5.1 & 561 \\
        49 & 1e-5  & -2124.078965 & 1.839 & 0.03 & 5.3 & 910 \\
        49 & 5e-6  & -2124.078976 & 1.828 & 0.04 & 5.3 & 912 \\
        49 & 1e-6  & -2124.079056 & 1.748 & 0.04 & 6.0 & 1034\\
        49 & 5e-7  & -2124.079150 & 1.654 & 0.05 & 6.1 & 1261\\
        49 & 1e-7  & -2124.079592 & 1.212 & 0.3 & 8 & 4945 \\
        49 & 5e-8  & -2124.079946 & 0.858 & 0.8 & 14 & 12950 \\
        49 & 1e-8  & -2124.080649 & 0.155 & 5 & 90 & 97618 \\
        49 & 5e-9  & -2124.080773 & 0.031 & 8 & 184 & 175359 \\
        49 & 1e-9  & \textbf{-2124.080804} & \boldmath$\coloneq 0$ & 11 & 401 & 354689 \\
        49 & 5e-10 & -2124.080797 & 0.007 & 12 & 413 & 366254 \\
        49 & 1e-10 & -2124.080778 & 0.026 & 16 & 494 & 448700 \\
        49 & 5e-11 & -2124.080757 & 0.047 & 25 & 679 & 584372 \\
        49 & 1e-11 & -2124.080718 & 0.086 & 144 & 3009 & 2152341 \\
        49 & 5e-12 & -2124.080707 & 0.097 & 348 & 9554 & 4500406 \\
        49 & 1e-12 & -2124.080678 & 0.126 & 1544 & 76787 & 20629402 \\
        49 & 5e-13 & -2124.080667 & 0.137 & 2373 & 117958 & 34122288 \\
        49 & 1e-13 & -2124.080656 & 0.148 & 3931 & 257395 & 71481681 \\
        49 & 5e-14 & -2124.080654 & 0.150 & 4262 & 302612 & 82373739 \\
        49 & 1e-14 & -2124.080652 & 0.152 & 7151 & 429161 & 112983846 \\
        \bottomrule \noalign{\vskip 1mm} 
    \end{tabular}
    \caption{Sparse simulator results for running the S$_0$ state of Ir(F$_2$ppy)$_3$ using the  LANL2DZ-ECP basis set with different amplitude cutoffs (lowest energy in bold and $\Delta$ is the difference to that energy). Amplitudes $\alpha_i$ are kept if $|\alpha_i|^2 \geq \text{cutoff}$. We used the optimized parameters $\beta^*$ from iQCC. Each iQCC step contains 20 QCC entanglers of the form \eqref{eq:qcc_entanglers}. For comparison, the energy obtained using the iQCC algorithm on a classical computer resulted in an energy of $-2124.081161$Ha.
    }
    \label{tab:s0_results_opt_amp}
\end{table*}

\begin{table*}[t]
    \centering
    \begin{tabular}{cclrrrr}
        \toprule
        \#iQCC steps & cutoff & energy [Ha] & $\Delta$ [mHa] & sim. time [s] & meas. time [s] & \#elements \\\midrule
        49 & 5e-5  & -2123.966249 & 16.732 & 0.03 & 5.1 & 515 \\
        49 & 1e-5  & -2123.981231 & 1.750 & 0.04 & 5.4 & 969 \\
        49 & 5e-6  & -2123.981312 & 1.669 & 0.04 & 5.4 & 979 \\
        49 & 1e-6  & -2123.981597 & 1.384 & 0.06 & 5.7 & 1334 \\
        49 & 5e-7  & -2123.981591 & 1.390 & 0.10 & 6.2 & 1952 \\
        49 & 1e-7  & -2123.981931 & 1.050 & 0.5 & 12 & 9110 \\
        49 & 5e-8  & -2123.982244 & 0.737 & 1.1 & 19 & 18737 \\
        49 & 1e-8  & -2123.982958 & 0.023 & 5  & 98  & 99527 \\
        49 & 5e-9  & \textbf{-2123.982981} & \boldmath$\coloneq 0$ & 7  & 196 & 171433 \\
        49 & 1e-9  & -2123.982918 & 0.063 & 13 & 487 & 388713 \\
        49 & 5e-10 & -2123.982909 & 0.072 & 14 & 530 & 428649 \\
        49 & 1e-10 & -2123.982871 & 0.110 & 28 & 825 & 640668 \\
        49 & 5e-11 & -2123.982856 & 0.125 & 45 & 1178 & 926128 \\
        49 & 1e-11 & -2123.982813 & 0.168 & 216 & 4953 & 3255603 \\
        49 & 1e-12 & -2123.982769 & 0.212 & 1601 & 86217 & 22094533 \\
        49 & 1e-13 & -2123.982745 & 0.236 & 4515 & 301017 & 80003357 \\
        \bottomrule \noalign{\vskip 1mm} 
    \end{tabular}
    \caption{Sparse simulator results for the T$_1$ state of Ir(F$_2$ppy)$_3$ using LANL2DZ-ECP basis set with different amplitude cutoffs (lowest energy in bold and $\Delta$ is the difference to that energy). Amplitudes $\alpha_i$ are kept if $| \alpha_i|^2 \geq \text{cutoff}$. We used the optimized parameters $\beta^*$ from iQCC. Each iQCC step contains 20 QCC entanglers of the form \eqref{eq:qcc_entanglers}. For comparison, the energy obtained using the iQCC algorithm on a classical computer resulted in an energy of $-2123.982805$Ha.
    }
    \label{tab:t1_results_opt_amp}
\end{table*}

\begin{figure}[ht]
    \includegraphics[width=\linewidth]{energy\_decrease\_ir\_s0.pdf}
    \caption{Energy decrease of each of the 49 iQCC steps for the S$_0$ state of Ir(F$_2$ppy)$_3$ using the LANL2DZ-ECP basis set. Total energy decrease of the 49 iQCC steps is $76.9$mHa. Each iQCC step contains 20 QCC entanglers of the form \eqref{eq:qcc_entanglers}. The simulator kept amplitudes if $| \alpha_i|^2 \geq 10^{-11}$. It can be seen that each set of QCC entanglers successfully lowers the overall energy.}
    \label{fig:energy_decrease_s0}
\end{figure}

\begin{figure}[ht]
    \includegraphics[width=\linewidth]{runtime\_vs\_nonzero\_elements.pdf}
    \caption{Sparse simulator measurement times as a function of the number of nonzero elements in the hash map. We used the data from the S$_0$ state of the Ir(F$_2$ppy)$_3$ simulation using the LANL2DZ-ECP basis set and varied the simulator amplitude cutoff to obtain different numbers of amplitudes in the hash map. The raw data is listed in Table~\ref{tab:s0_results_opt_amp}. As expected, we observe a linear scaling.}
    \label{fig:runtime_vs_nonzero_elements}
\end{figure}

\begin{table*}[t]
    \centering
    \begin{tabular}{cclrrr}
        \toprule
        \#iQCC steps & cutoff & energy [Ha] & sim. time [s] & meas. time [s] & \#elements \\\midrule
        55  & 1e-12 & -2124.225246 & 1417 & 54654 & 16349362 \\
        \bottomrule \noalign{\vskip 1mm} 
    \end{tabular}
    \caption{Sparse simulator results for running the S$_0$ state of Ir(F$_2$ppy)$_3$ using the SBKJC-ECP  basis set. Amplitudes $\alpha_i$ are kept if $| \alpha_i|^2 \geq \text{cutoff}$. We used the optimized parameters $\beta^*$ from iQCC. Each iQCC step contains 20 QCC entanglers of the form \eqref{eq:qcc_entanglers}. For comparison, the energy obtained using the iQCC algorithm on a classical computer resulted in an energy of $-2124.226942$Ha.
}
    \label{tab:s0_sbkjc_results_opt_amp}
\end{table*}

\subsection{Sparsity analysis}
\begin{figure}[ht]
    \includegraphics[width=\linewidth]{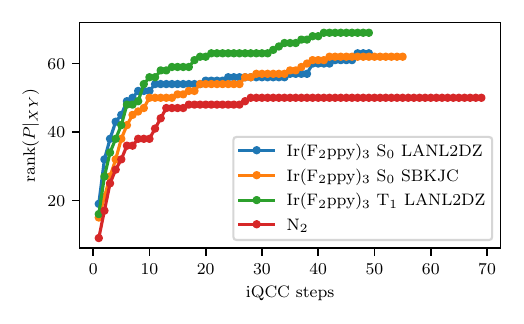}
    \caption{For all systems we compute an upper bound on the number of nonzero elements in the wave function as $2^{\mathrm{rank}(P|_{XY})}$ for a given iQCC ansatz. Each iQCC step contains 20 entanglers of the form \eqref{eq:qcc_entanglers}. The Ir(F$_2$ppy)$_3$ system has 80 qubits, while the $\texttt{N}_2$ system has 56 qubits. }
    \label{fig:ranks}
\end{figure}

We analysed the upper bound on the number of nonzero elements in the trial wave functions, see Fig.~\ref{fig:ranks}. In all cases for the complete set of entanglers, it would not be feasible to store all nonzero elements of the wave function. However, we empirically found that truncation can be used to make storing the wave function feasible without having to sacrifice too much accuracy; see Fig.~\ref{fig:cutoff}. Of course, this does not completely rule out the possibility that there might be significant effects due to such a truncation. Investigating this would be an interesting benchmark for future quantum computers.

\subsection{Quantum resources}\label{sec:sub_quantum_resources}
Unfortunately, current quantum computing hardware is unable to run the circuits used in this work. However, we estimate the number of gates required to do so. We used ProjectQ \cite{steiger2018projectq} to compile the circuit and perform some basic optimisations. The compiler assumes that qubits feature all-to-all connectivity (for CNOT gates), which for some quantum hardware would imply a further mapping overhead. Our estimates can be found in Table~\ref{tab:qc_resources}.

\begin{table*}[t]
    \centering
    \begin{tabular}{lccccccrr}
        \toprule
        Molecule & \#iQCC steps & \#Qubits&\#CNOT & \#X & \#H & \#$R_x$ & \#$R_z$ & \#terms in $H$ \\\midrule
        Ir(F$_2$ppy)$_3$ S$_0$ state (LANL2DZ) & 49 & 80 & 5420 & 40 & 776 & 732 & 976 & 3775249 \\
        Ir(F$_2$ppy)$_3$ T$_1$ state (LANL2DZ) & 49 & 80 & 5436 & 40 & 728 & 672 & 980 & 3774557 \\
        Ir(F$_2$ppy)$_3$ S$_0$ state (SBKJC) & 55 & 80 & 6344 & 40 & 818 & 756 & 1097 & 3676229\\
        N$_2$ & 69 & 56 & 8252 & 12 & 934 & 850 & 1379 & 190753 \\
        \bottomrule \noalign{\vskip 1mm} 
    \end{tabular}
    \caption{Number of quantum resources to implement the iQCC circuit. Each iQCC step contains 20 entanglers of the form \eqref{eq:qcc_entanglers}.}
    \label{tab:qc_resources}
\end{table*}

\section{Discussion}
We have demonstrated that sparse wave function simulation can be used to simulate large VQE circuits generated by iQCC. The majority of the running time of our code is spent in the measurement phase, for which we thus decided to use multiple threads. As expected, our code showed excellent strong scaling behavior up to 192 threads. Compared to iQCC, we observed that sparse simulation required two orders of magnitude less memory for our benchmarks.
If one is interested in computing energy lower bounds classically instead of requiring a converged quantum circuit simulation, then even more substantial memory savings are possible, e.g., an additional $58\times$ for the $S_0$ state of the Ir(F$_2$ppy)$_3$ system.
We found that all our simulation results were in good agreement with iQCC values and, where available, with experimental results.

Our sparse simulator has two additional advantages compared to iQCC, stemming from the fact that it measures the original Hamiltonian (as opposed to an approximate, similarity-transformed version) one term at a time. First, this allows us to compare to measurements of future NISQ computers, which also measure the expectation value of each term separately. Second, by using the original Hamiltonian, we ensure that our results are upper bounds on the exact energy. We saw that substantial differences may be observed when using methods without such a guarantee, e.g., for the test case of N$_2$, where there is an energy difference of 3.5mHa between the energy reported by iQCC and the converged energy reported by our simulator. This is likely due to the accumulation of truncation errors in iQCC that lead to a final Hamiltonian that is no longer isospectral to the original Hamiltonian.

Our analysis of the maximum number of nonzero amplitudes in subsection~\ref{sec:sparsity_analysis} also provides a constructive approach to build VQE circuits that may be simulated effectively and/or even exactly using our approach. This is useful to produce test instances for NISQ devices with 50-100 qubits and up to thousands of entanglers of the form given in \eqref{eq:pauli_exponentials}. As in iQCC, one can choose entanglers that have a large gradient at parameter value $\beta_i=0$ with the additional constraint that these new entanglers have Pauli operators $P_i$ such that the rank of the $P|_{XY}$ matrix does not increase. This can be accomplished by only allowing new Pauli operators $P_i$ which have an associated $P_i|_{XY}$ that is a linear combination of previous $P_{j<i}|_{XY}$ masks.

We also envision our sparse simulator to work in tandem with iQCC, where iQCC produces a suitable set of entanglers as a starting point for further parameter optimisation using our sparse simulator.

\printbibliography

\end{document}

%% file: time_evolution.tikz
\providecommand{\ket}[1]{\left|#1\right\rangle}
\begin{tikzpicture}[scale=1.000000,x=1pt,y=1pt]
\filldraw[color=white] (0.000000, -7.500000) rectangle (319.000000, 37.500000);
\draw[color=black] (0.000000,30.000000) -- (319.000000,30.000000);
\draw[color=black] (0.000000,30.000000) node[left] {$q_0$};
\draw[color=black] (0.000000,15.000000) -- (319.000000,15.000000);
\draw[color=black] (0.000000,15.000000) node[left] {$q_1$};
\draw[color=black] (0.000000,0.000000) -- (319.000000,0.000000);
\draw[color=black] (0.000000,0.000000) node[left] {$q_2$};
\draw (12.000000, -7.500000) node[text width=144pt,below,text centered] {\color{blue}\textcircled{\scriptsize{0}}};
\begin{scope}
\draw[fill=white] (35.500000, 30.000000) +(-45.000000:8.485281pt and 8.485281pt) -- +(45.000000:8.485281pt and 8.485281pt) -- +(135.000000:8.485281pt and 8.485281pt) -- +(225.000000:8.485281pt and 8.485281pt) -- cycle;
\clip (35.500000, 30.000000) +(-45.000000:8.485281pt and 8.485281pt) -- +(45.000000:8.485281pt and 8.485281pt) -- +(135.000000:8.485281pt and 8.485281pt) -- +(225.000000:8.485281pt and 8.485281pt) -- cycle;
\draw (35.500000, 30.000000) node {$H$};
\end{scope}
\draw (35.500000, -7.500000) node[text width=144pt,below,text centered] {\hspace{47pt}\color{blue}\textcircled{\scriptsize{1}}};
\begin{scope}
\draw[fill=white] (35.500000, 15.000000) +(-45.000000:24.748737pt and 10.606602pt) -- +(45.000000:24.748737pt and 10.606602pt) -- +(135.000000:24.748737pt and 10.606602pt) -- +(225.000000:24.748737pt and 10.606602pt) -- cycle;
\clip (35.500000, 15.000000) +(-45.000000:24.748737pt and 10.606602pt) -- +(45.000000:24.748737pt and 10.606602pt) -- +(135.000000:24.748737pt and 10.606602pt) -- +(225.000000:24.748737pt and 10.606602pt) -- cycle;
\draw (35.500000, 15.000000) node {$R_x(\frac{\pi}{2})$};
\end{scope}
\draw (68.000000,30.000000) -- (68.000000,15.000000);
\begin{scope}
\draw[fill=white] (68.000000, 15.000000) circle(3.000000pt);
\clip (68.000000, 15.000000) circle(3.000000pt);
\draw (65.000000, 15.000000) -- (71.000000, 15.000000);
\draw (68.000000, 12.000000) -- (68.000000, 18.000000);
\end{scope}
\filldraw (68.000000, 30.000000) circle(1.500000pt);
\draw (86.000000, -7.500000) node[text width=144pt,below,text centered] {\hspace{18pt}\color{blue}\textcircled{\scriptsize{2}}};
\draw (86.000000,15.000000) -- (86.000000,0.000000);
\begin{scope}
\draw[fill=white] (86.000000, 0.000000) circle(3.000000pt);
\clip (86.000000, 0.000000) circle(3.000000pt);
\draw (83.000000, 0.000000) -- (89.000000, 0.000000);
\draw (86.000000, -3.000000) -- (86.000000, 3.000000);
\end{scope}
\filldraw (86.000000, 15.000000) circle(1.500000pt);
\draw (118.500000, -7.500000) node[text width=144pt,below,text centered] {\hspace{47pt}\color{blue}\textcircled{\scriptsize{3}}};
\begin{scope}
\draw[fill=white] (118.500000, -0.000000) +(-45.000000:24.748737pt and 10.606602pt) -- +(45.000000:24.748737pt and 10.606602pt) -- +(135.000000:24.748737pt and 10.606602pt) -- +(225.000000:24.748737pt and 10.606602pt) -- cycle;
\clip (118.500000, -0.000000) +(-45.000000:24.748737pt and 10.606602pt) -- +(45.000000:24.748737pt and 10.606602pt) -- +(135.000000:24.748737pt and 10.606602pt) -- +(225.000000:24.748737pt and 10.606602pt) -- cycle;
\draw (118.500000, -0.000000) node {$R_z(2\theta)$};
\end{scope}
\draw (151.000000,15.000000) -- (151.000000,0.000000);
\begin{scope}
\draw[fill=white] (151.000000, 0.000000) circle(3.000000pt);
\clip (151.000000, 0.000000) circle(3.000000pt);
\draw (148.000000, 0.000000) -- (154.000000, 0.000000);
\draw (151.000000, -3.000000) -- (151.000000, 3.000000);
\end{scope}
\filldraw (151.000000, 15.000000) circle(1.500000pt);
\draw (169.000000, -7.500000) node[text width=144pt,below,text centered] {\hspace{18pt}\color{blue}\textcircled{\scriptsize{4}}};
\draw (169.000000,30.000000) -- (169.000000,15.000000);
\begin{scope}
\draw[fill=white] (169.000000, 15.000000) circle(3.000000pt);
\clip (169.000000, 15.000000) circle(3.000000pt);
\draw (166.000000, 15.000000) -- (172.000000, 15.000000);
\draw (169.000000, 12.000000) -- (169.000000, 18.000000);
\end{scope}
\filldraw (169.000000, 30.000000) circle(1.500000pt);
\begin{scope}
\draw[fill=white] (201.500000, 30.000000) +(-45.000000:8.485281pt and 8.485281pt) -- +(45.000000:8.485281pt and 8.485281pt) -- +(135.000000:8.485281pt and 8.485281pt) -- +(225.000000:8.485281pt and 8.485281pt) -- cycle;
\clip (201.500000, 30.000000) +(-45.000000:8.485281pt and 8.485281pt) -- +(45.000000:8.485281pt and 8.485281pt) -- +(135.000000:8.485281pt and 8.485281pt) -- +(225.000000:8.485281pt and 8.485281pt) -- cycle;
\draw (201.500000, 30.000000) node {$H$};
\end{scope}
\draw (201.500000, -7.500000) node[text width=144pt,below,text centered] {\hspace{47pt}\color{blue}\textcircled{\scriptsize{5}}};
\begin{scope}
\draw[fill=white] (201.500000, 15.000000) +(-45.000000:24.748737pt and 10.606602pt) -- +(45.000000:24.748737pt and 10.606602pt) -- +(135.000000:24.748737pt and 10.606602pt) -- +(225.000000:24.748737pt and 10.606602pt) -- cycle;
\clip (201.500000, 15.000000) +(-45.000000:24.748737pt and 10.606602pt) -- +(45.000000:24.748737pt and 10.606602pt) -- +(135.000000:24.748737pt and 10.606602pt) -- +(225.000000:24.748737pt and 10.606602pt) -- cycle;
\draw (201.500000, 15.000000) node {$R_x(\frac{-\pi}{2})$};
\end{scope}
\draw[fill=white,color=white] (231.000000, -6.000000) rectangle (246.000000, 36.000000);
\draw (238.500000, 15.000000) node {$=$};
\draw (285.500000,30.000000) -- (285.500000,0.000000);
\begin{scope}
\draw[fill=white] (285.500000, 15.000000) +(-45.000000:38.890873pt and 29.698485pt) -- +(45.000000:38.890873pt and 29.698485pt) -- +(135.000000:38.890873pt and 29.698485pt) -- +(225.000000:38.890873pt and 29.698485pt) -- cycle;
\clip (285.500000, 15.000000) +(-45.000000:38.890873pt and 29.698485pt) -- +(45.000000:38.890873pt and 29.698485pt) -- +(135.000000:38.890873pt and 29.698485pt) -- +(225.000000:38.890873pt and 29.698485pt) -- cycle;
\draw (285.500000, 15.000000) node {$e^{-i \theta X_0 Y_1 Z_2}$};
\end{scope}
\draw[dashed] (12.000000, -7.500000) -- (12.000000, 37.500000);
\draw[dashed] (59.000000, -7.500000) -- (59.000000, 37.500000);
\draw[dashed] (95.000000, -7.500000) -- (95.000000, 37.500000);
\draw[dashed] (142.000000, -7.500000) -- (142.000000, 37.500000);
\draw[dashed] (178.000000, -7.500000) -- (178.000000, 37.500000);
\draw[dashed] (225.000000, -7.500000) -- (225.000000, 37.500000);
\end{tikzpicture}